\documentclass[twocolumn,english,aps,prl,showpacs]{revtex4}
\usepackage[T1]{fontenc}
\usepackage[latin9]{inputenc}
\usepackage{amsmath}
\usepackage{amssymb}
\usepackage{graphicx}
\usepackage{esint}

\makeatletter
\@ifundefined{textcolor}{}
{%
 \definecolor{BLACK}{gray}{0}
 \definecolor{WHITE}{gray}{1}
 \definecolor{RED}{rgb}{1,0,0}
 \definecolor{GREEN}{rgb}{0,1,0}
 \definecolor{BLUE}{rgb}{0,0,1}
 \definecolor{CYAN}{cmyk}{1,0,0,0}
 \definecolor{MAGENTA}{cmyk}{0,1,0,0}
 \definecolor{YELLOW}{cmyk}{0,0,1,0}
}

\makeatother

\makeatother

\usepackage{babel}
\begin{document}

\title{First and second sound in a two-dimensional dilute Bose gas across
the Berezinskii-Kosterlitz-Thouless transition}

\author{Xia-Ji Liu$^{1}$ and Hui Hu$^{1}$ }

\affiliation{$^{1}$Centre for Quantum and Optical Science, Swinburne University
of Technology, Melbourne 3122, Australia}

\date{\today}
\begin{abstract}
We theoretically investigate first and second sound of a two-dimensional
(2D) atomic Bose gas in harmonic traps by solving Landau's two-fluid
hydrodynamic equations. For an isotropic trap, we find that first
and second sound modes become degenerate at certain temperatures and
exhibit typical avoided crossings in mode frequencies. At these temperatures,
second sound has significant density fluctuation due to its hybridization
with first sound and has a divergent mode frequency towards the Berezinskii-Kosterlitz-Thouless
(BKT) transition. For a highly anisotropic trap, we derive the simplified
one-dimensional hydrodynamic equations and discuss the sound-wave
propagation along the weakly confined direction. Due to the universal
jump of the superfluid density inherent to the BKT transition, we
show that the first sound velocity exhibits a kink across the transition.
Our predictions can be readily examined in current experimental setups
for 2D dilute Bose gases.
\end{abstract}

\pacs{67.85.De, 03.75.Kk, 05.30.Jp}

\maketitle
Low-energy excitations of a quantum liquid in its superfluid state
- in which inter-particle collisions are sufficiently frequent to
ensure local thermodynamic equilibrium - can be well described by
Landau's two-fluid hydrodynamic theory \cite{Tisza1938,Landau1941}.
It is now widely known that there are two types of excitations, namely
first and second sound, which describe respectively the coupled in-phase
(density) and out-of-phase (temperature) oscillations of the superfluid
and normal fluid components \cite{GriffinBook}. Historically, Landau's
two-fluid hydrodynamic theory was invented to understand the quantum
liquid of superfluid helium \cite{Landau1941}. The study of first
and second sound in such a system has greatly enriched our knowledge
of the fascinating but challenging many-body physics. For any new
kind quantum fluids, it is therefore natural to anticipate that first
and second sound may also provide a powerful tool to characterize
their underlying physics.

\begin{figure}
\begin{centering}
\includegraphics[clip,width=0.45\textwidth]{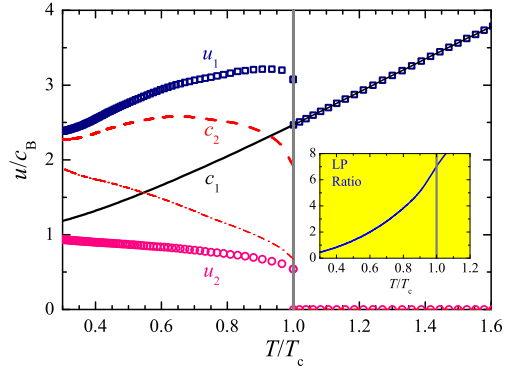} 
\par\end{centering}

\caption{(Color online) First (squares) and second sound velocities (circles)
of a uniform 2D Bose gas across the BKT transition temperature $T_{c}$,
in unit of the Bogoliubov sound velocity $c_{B}$. The dimensionless
coupling constant $g=0.05$. For comparison, the decoupled first and
second sound velocities are shown by the solid and dashed lines, respectively.
The dot-dashed line is the velocity of second sound with the leading-order
correction due to its coupling to first sound. The inset shows the
Landau-Placzek parameter, which characterizes the coupling between
first and second sound.}

\label{fig1} 
\end{figure}

In this context, the case of a two-dimensional (2D) dilute Bose gas
confined in harmonic trapping potentials, which has recently been
realized in ultracold atomic laboratory \cite{Hung2011,Yefsah2011,Desbuquois2012,Ha2013},
is of particular interest. At nonzero temperatures, the condensation
of bosonic atoms is precluded by the Hohenberg-Mermin-Wagener theorem
\cite{Hohenberg1967,Mermin1966}. The superfluid phase transition
in such a system is of the Berezinskii-Kosterlitz-Thouless (BKT) type
\cite{Berezinskii1972,Kosterlitz1972}, whose nature is remarkably
different from the conventional second-order phase transition in three
dimensions (3D). The BKT transition is associated with the emergence
of a topological order, as a result of the pairing of vortices and
anti-vortices. Therefore, across the BKT transition from below, the
superfluid density of the system jumps to zero from a universal value
$4/\lambda_{dB}^{2}$, where $\lambda_{dB}\equiv\sqrt{2\pi\hbar^{2}/(mk_{B}T)}$
is the thermal de Broglie wavelength at the temperature $T$. In the
absence of harmonic traps, this leads to discontinuities in the first
and second sound velocities at the BKT transition \cite{Ozawa2014},
as illustrated in Fig. \ref{fig1}.

In this work, we discuss the behavior of first and second sound of
a harmonically \emph{trapped} 2D Bose gas. Our investigation is motivated
by the recent sound mode measurements in a trapped 3D unitary Fermi
gas \cite{Altmeyer2007,Joseph2007,Tey2013,Sidorenkov2013}, which
provide valuable information on its equation of state and superfluid
density. In particular, in a milestone experiment performed by the
Innsbruck team \cite{Sidorenkov2013}, first and second sound waves
were excited in the highly elongated unitary Fermi gas, and their
propagations along the weakly confined axis were measured. These measurements
are straightforward to implement in a trapped 2D Bose or Fermi gas,
and would greatly promote the current experimental \cite{Hung2011,Yefsah2011,Desbuquois2012,Ha2013}
and theoretical research \cite{Prokofev2001,Prokofev2002,Rancon2012}
on the intriguing BKT physics in ultracold atoms and solid-state systems.

Our main results are briefly summarized as follows. In sharp contrast
to the superfluid helium and unitary Fermi gas, we find that the coupling
between density and temperature oscillations in a 2D Bose gas are
very strong, as characterized by a large Landau-Placzek (LP) parameter
(see the inset of Fig. \ref{fig1}), thereby making the observation
of second sound much easier. The universal jump of superfluid density
at the BKT transition leads to non-trivial consequences in the sound
mode frequencies and velocities. In an isotropic harmonic trap, we
show that the mode frequency of second sounds diverges as $(T-T_{c})^{-1/2}$
approaching the BKT critical temperature $T_{c}$. While in a highly
anisotropic trap, the velocity of the first sound wave propagation
in the weakly confined direction exhibits an apparent kink right at
the transition. The experimental confirmation of these predictions
would provide a complete proof of the BKT physics.

We start by considering an interacting atomic Bose gas trapped in
a 2D harmonic potential, $V_{T}(\mathbf{r})=m(\omega_{x}^{2}x^{2}+\omega_{y}^{2}y^{2})/2,$
with atomic mass $m$ and trapping frequencies $\omega_{x}$ and $\omega_{y}$.
The motion in the third direction is assumed to be frozen by an additional,
tight harmonic confinement, as realized in current experiments \cite{Hung2011,Yefsah2011,Desbuquois2012,Ha2013}.
First and second sound of the system are described by Landau's two-fluid
hydrodynamic equations which involve only local thermodynamic variables
and superfluid density. As discussed in the previous works \cite{Taylor2005,Taylor2008,Taylor2009},
using Hamilton's variational principle \cite{Zilsel1950}, first and
second sound modes of these equations with frequency $\omega$ can
be obtained by minimizing a variational action, which, in terms of
displacement fields $\mathbf{u}_{s}(\mathbf{r})$ and $\mathbf{u}_{n}(\mathbf{r})$
for superfluid and normal fluid components, takes the following form
\cite{Taylor2005},
\begin{eqnarray}
S & = & \frac{1}{2}\int d\mathbf{r}\left[m\omega^{2}\left(n_{s}\mathbf{u}_{s}^{2}+n_{n}\mathbf{u}_{n}^{2}\right)-\left(\frac{\partial\mu}{\partial n}\right)_{s}\left(\delta n\right)^{2}\right.\nonumber \\
 &  & \left.-2\left(\frac{\partial T}{\partial n}\right)_{s}\delta n\delta s-\left(\frac{\partial T}{\partial s}\right)_{n}\left(\delta s\right)^{2}\right].\label{eq:usunAction}
\end{eqnarray}
Here, $n(\mathbf{r})$ and $s(\mathbf{r})$ are respectively the local
number density and entropy density, $n_{s}(\mathbf{r})$ and $n_{n}(\boldsymbol{\mathbf{r}})=n(\mathbf{r})-n_{s}(\mathbf{r})$
are the superfluid and normal-fluid densities at equilibrium, $\delta n\mathbf{(r})\equiv-\mathbf{\nabla}\cdot(n_{s}\mathbf{u}_{s}+n_{n}\mathbf{u}_{n})$
is the density fluctuation, and $\delta s(\mathbf{r})\equiv-\mathbf{\nabla}\cdot(s\mathbf{u}{}_{n}$)
is the entropy fluctuation. The effect of the harmonic trapping potential
$V_{T}(\mathbf{r})$ enters the action Eq. (\ref{eq:usunAction})
through the coordinate dependence of the equilibrium thermodynamic
variables $(\partial\mu/\partial n)_{s}$, $(\partial T/\partial n)_{s}$
and $(\partial T/\partial s)_{n}$. 

For a 2D interacting Bose gas, due to the scale invariance of the
interatomic interaction \cite{Pitaevskii1997}, all the thermodynamic
inputs can be written in terms of dimensionless universal functions
that depend only on the ratio $z(\mathbf{r})=\mu(\mathbf{r})/k_{B}T$
and the dimensionless coupling constant $g=\sqrt{8\pi}a_{s}/l_{z}$
\cite{Prokofev2001}, where $\mu(\mathbf{r})=\mu-V_{T}(\mathbf{r})$
is the local chemical potential within local density approximation,
$a_{s}$ is the 3D scattering length, and $l_{z}$ is the oscillator
length in the tight confinement direction. In particular, the local
pressure, number density and superfluid density are given by \cite{Ozawa2014},
$P(\mathbf{r})=k_{B}T\lambda_{dB}^{-2}f_{p}[g,z(\mathbf{r})]$, $n(\mathbf{r})=\lambda_{dB}^{-2}f_{n}[g,z(\mathbf{r})]$
and $n_{s}(\mathbf{r})=\lambda_{dB}^{-2}f_{s}[g,z(\mathbf{r})]$,
respectively. These universal functions have been calculated theoretically
\cite{Prokofev2001,Prokofev2002,Rancon2012} and partly measured experimentally
\cite{Hung2011,Yefsah2011}, to certain accuracy. Throughout this
work, we will use the results determined by Monte Carlo simulations
for small interaction parameter $g$ \cite{Prokofev2001,Prokofev2002}. 

In superfluid helium \cite{GriffinBook} and unitary Fermi gas \cite{Taylor2009},
the solutions of Landau's hydrodynamic equations can be well understood
as density and temperature waves, which are the pure in-phase mode
with $\mathbf{u}_{s}=\mathbf{u}_{n}$ and the pure out-of-phase mode
with $n_{s}\mathbf{u}_{s}+n_{n}\mathbf{u}_{n}=0$, known as first
and second sound, respectively \cite{Landau1941,GriffinBook}. Following
this classification, we may rewrite the action Eq. (\ref{eq:usunAction})
in terms of two new displacement fields $\mathbf{u}_{a}=(n_{s}\mathbf{u}_{s}+n_{n}\mathbf{u}_{n})/n$
and $\mathbf{u}_{e}=\mathbf{u}_{s}-\mathbf{u}_{n},$ since the density
fluctuation $\delta n=-\mathbf{\nabla}\cdot(n\mathbf{u}_{a})$ and
the temperature fluctuation is given by $\delta T=(\partial T/\partial s)_{n}\mathbf{\nabla}\cdot(sn_{s}\mathbf{u}_{e}/n)$.
Roughly speaking, first sound is characterized by $\delta n\neq0$
but $\delta T=0$ and second sound by $\delta n=0$ but $\delta T\neq0$.
Using the standard thermodynamic identities, after some straightforward
but lengthy algebra, we arrive at $S=(1/2)\int d\mathbf{r}[\mathcal{S}^{(a)}+2\mathcal{S}^{(ae)}+\mathcal{S}^{(e)}]$,
where
\begin{eqnarray}
\mathcal{S}^{(a)} & = & m\omega^{2}n\mathbf{u}_{a}^{2}-n\left(\frac{\partial P}{\partial n}\right)_{\bar{s}}\left(\mathbf{\nabla}\cdot\mathbf{u}_{a}\right)^{2}+\mathcal{S}^{\left(V\right)},\label{eq:S(a)}\\
\mathcal{S}^{(ae)} & = & \left(\frac{\partial P}{\partial s}\right)_{n}\left(\mathbf{\nabla}\cdot\mathbf{u}_{a}\right)\left[\mathbf{\nabla}\cdot\left(\frac{sn_{s}}{n}\mathbf{u}_{e}\right)\right],\label{eq:S(ae)}\\
\mathcal{S}^{(e)} & = & m\omega^{2}\frac{n_{s}n_{n}}{n}\mathbf{u}_{e}^{2}-\left(\frac{\partial T}{\partial s}\right)_{n}\left[\mathbf{\nabla}\cdot\left(\frac{sn_{s}}{n}\mathbf{u}_{e}\right)\right]^{2},\label{eq:S(e)}
\end{eqnarray}
and $\mathcal{S}^{(V)}\equiv(\nabla n\cdot\mathbf{u}_{a})(\mathbf{\nabla}V_{T}\cdot\mathbf{u}_{a})+2n(\mathbf{\nabla}V_{T}\cdot\mathbf{u}_{a})(\nabla\cdot\mathbf{u}_{a})$
is the part directly related to the trapping potential $V_{T}$ and
$\bar{s}=s/n$ is the entropy per particle. It is clear that the first
and second sound are governed by the actions $\mathcal{S}^{(a)}$
and $\mathcal{S}^{(e)}$, respectively. Their coupling is controlled
by the term $\mathcal{S}^{(ae)}$, which generally is nonzero. In
our case of a 2D Bose gas, the scale invariance leads to $(\partial P/\partial s)_{n}=T$,
indicating that first and second sound are coupled at any nonzero
temperature.

For a uniform superfluid ($V_{T}=0$), the solutions of $\mathcal{S}^{(a)}$
and $\mathcal{S}^{(e)}$ are plane waves of wave vector $q$ with
dispersion $\omega_{1}=c_{1}q$ and $\omega_{2}=c_{2}q$, where $c_{1}=\sqrt{\left(\partial P/\partial n\right)_{\bar{s}}/m}$,
$c_{2}=\sqrt{k_{B}T\bar{s}^{2}n_{s}/(m\bar{c}_{v}n_{n})}$, and $\bar{c}_{v}$
is the specific heat per particle at constant volume. In Fig. \ref{fig1},
we report the temperature dependence of the decoupled first and second
sound velocities at the coupling constant $g=0.05$ by the black solid
and red dashed lines, respectively, measured in unit of the zero temperature
Bogoliubov sound velocity $c_{B}=\hbar\sqrt{gn}/m$. At the BKT transition
temperature $T_{c}=2\pi\hbar^{2}n/[mk_{B}\ln(\xi/g)]$, where $\xi=380\pm3$
is a universal parameter \cite{Prokofev2001,Prokofev2002}, the decoupled
second sound velocity exhibits a discontinuity due to the universal
jump in superfluid density. Including the coupling term $\mathcal{S}^{(ae)}$,
we obtain the standard hydrodynamic equation for sound velocity $u$
\cite{Landau1941,GriffinBook}:
\begin{equation}
u^{4}-u^{2}\left(c_{1}^{2}+c_{2}^{2}\right)+c_{1}^{2}c_{2}^{2}/\gamma=0,\label{eq:soundvelocity}
\end{equation}
where $\gamma\equiv\bar{c}_{p}/\bar{c}_{v}$ is the ratio between
specific heats at const pressure and volume. The coupling between
first and second sound can be conveniently characterized by the so-called
LP parameter $\epsilon_{\textrm{LP}}=\gamma-1$ \cite{Taylor2009,Hu2010}.
There are two solutions for the above hydrodynamic equation, $u_{1}$
and $u_{2}$, which in the absence of $\mathcal{S}^{(ae)}$ (i.e.,
$\gamma=1$ or $\epsilon_{\textrm{LP}}=0$), coincide with the decoupled
first and second sound velocities, $c_{1}$ and $c_{2}$. In Fig.
\ref{fig1}, we present the velocities $u_{1}$ and $u_{2}$ by squares
and circles, respectively. It is not a surprise to see that both velocities
shows discontinuity at the BKT transition, as discussed in Ref. \cite{Ozawa2014}.
Remarkably, the sound velocities $u_{1}$ and $u_{2}$ differ largely
from their decoupled counterparts $c_{1}$ and $c_{2}$, due to the
large value of the LP parameter (see the inset). This is in sharp
contrast to the cases of superfluid helium and unitary Fermi gas,
where $\epsilon_{\textrm{LP}}\sim0$ and hence first and second sound
couple very weakly. Nevertheless, near the transition, as shown by
the dot-dashed line in Fig. \ref{fig1}, we find that the second sound
velocity can still be approximated by $u_{2}\simeq c_{2}/\sqrt{\gamma}=\sqrt{k_{B}T\bar{s}^{2}n_{s}/(m\bar{c}_{p}n_{n})}$
\cite{Hu2010}, indicating that the second sound could be well regarded
a temperature wave at constant pressure. 

The strong coupling between first and second sound is of great importance
from the experimental point of view. In cold-atom experiments, temperature
oscillations can not be directly measured. Thus, the characterization
of second sound has to rely on the density measurement \cite{Sidorenkov2013}.
The strong coupling implies a large density fluctuation for second
sound and thus makes its observation much easier. Indeed, at the constant
pressure we find that the the ratio between the relative density and
temperature fluctuations is given by, 
\begin{equation}
\frac{\delta n/n}{\delta T/T}\simeq\frac{T}{n}\left(\frac{\partial n}{\partial T}\right)_{P}=-\epsilon_{\textrm{LP}},
\end{equation}
as shown in the inset of Fig. \ref{fig3}(a). A large LP parameter
therefore guarantees a significant density fluctuation of second sound
for experimental observation.

We now consider the experimentally relevant harmonic traps. Focusing
on an isotropic trapping potential ($\omega_{x}=\omega_{y}=\omega_{\perp}$)
and compressional breathing modes (i.e., angular momentum $l=0$),
we may solve the variational action by inserting the following polynomial
ansatz for the displacement fields \cite{Taylor2009}:
\begin{equation}
\mathbf{u}_{a}=\mathbf{\hat{r}}\sum_{i=0}^{N_{p}-1}A_{i}r^{i+1},\begin{array}{cc}
\end{array}\mathbf{u}_{e}=\left[\frac{n\left(r\right)}{n_{s}\left(r\right)}\right]\mathbf{\hat{r}}\sum_{i=0}^{N_{p}-1}B_{i}r^{i+1},
\end{equation}
where $\mathbf{\hat{r}}$ is the unit vector along the radial direction,
$\{A_{i},B_{i}\}$ ($i=0,\cdots,N_{p}-1$) are the $2N_{p}$ variational
parameters. The breathing mode frequencies are obtained by minimizing
the action $S$ with respect to these $2N_{p}$ parameters.

\begin{figure}
\begin{centering}
\includegraphics[clip,width=0.48\textwidth]{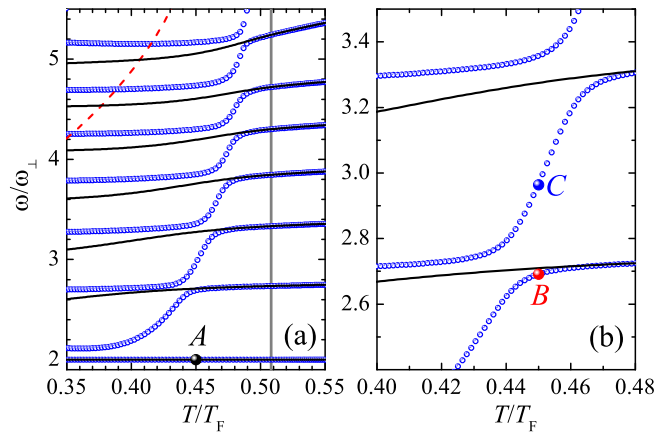} 
\par\end{centering}

\caption{(Color online) (a) Breathing mode frequencies of a 2D Bose gas with
$g=0.05$ trapped in an isotropic harmoinc potential of frequency
$\omega_{\perp}$. The full solutions calculated with $2N_{p}=16$
variational parameters are shown by blue symbols, and the decoupled
first and second sound solutions, $\omega_{1}$ and $\omega_{2}$,
are plotted by solid and dashed lines, respectively. (b) A enlarged
view of the left panel showing the hybridization between second sound
and the $n=1$ and $n=2$ first sound modes. The density fluctuations
at the points $A$, $B$ and $C$ are shown in Fig. \ref{fig3}(b).
The temperature is measured in unit of a Fermi temperature $T_{F}=(2N)^{1/2}\hbar\omega_{\perp}/k_{B}$
of a 2D ideal spinless Fermi gas with the same number of atoms $N$
as the Bose gas. The vertical gray line in (a) indicates the BKT transition
temperature in traps. It should be note that the lowest $n=0$ breathing
mode with frequency $2\omega_{\perp}$ is an exact solution of Landau's
two-fluid hydrodynamic equations \cite{Taylor2008}.}

\label{fig2} 
\end{figure}

In Fig. \ref{fig2}(a), we show the discretized mode frequencies of
the full two-fluid hydrodynamic action (blue symbols), as well as
the decoupled first and second sound mode frequencies determined by
$\mathcal{S}^{(a)}$ and $\mathcal{S}^{(e)}$ individually (black
solid and red dashed lines, respectively). The decoupled sound mode
frequencies differ largely from the full solutions, similar to the
uniform case. Despite the large difference, we may still classify
the first and second sound solutions as the horizontal and vertical
branches, respectively. These solutions become degenerate at certain
temperatures and hence exhibit clear avoided crossings with a typical
distance $\Delta\omega\sim0.2\omega_{\perp}$, as seen in the enlarged
view of Fig. \ref{fig2}(b). We note that, for a unitary Fermi gas,
similar avoided crossings have been predicted \cite{Taylor2009}.
However, their structure (i.e., $\Delta\omega\sim0.01\omega_{\perp}$)
seems to be too small to observe experimentally.  

It is evident that the second sound mode frequencies diverge towards
the BKT transition. This peculiar behavior is caused by the universal
jump in superfluid density. Approaching to the critical temperature
$T_{c}$, the size $R_{s}$ of the superfluid component decreases
as $R_{s}\propto\sqrt{T-T_{c}}$, leading to an increase in the minimum
wave vector, $q\sim1/R_{s}$. Using a finite second sound velocity
$c_{2}$ at the transition, we find that $\omega_{2}\sim c_{2}q\propto(T-T_{c})^{-1/2}$
and therefore a divergent second sound mode frequency.

\begin{figure}
\begin{centering}
\includegraphics[clip,width=0.45\textwidth]{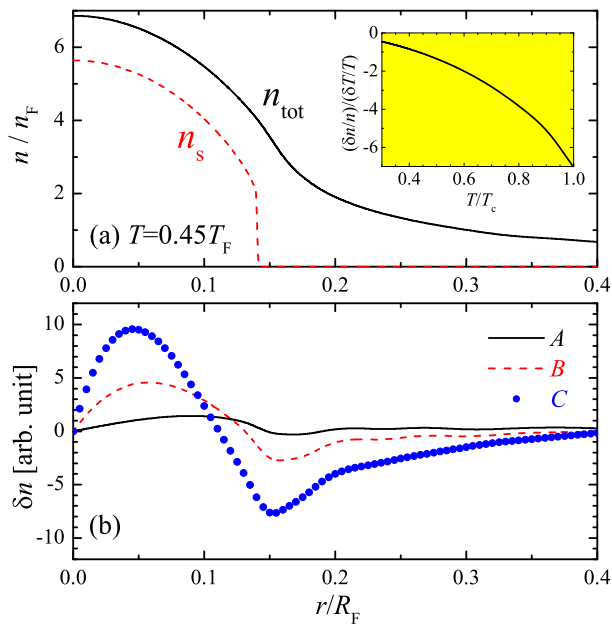} 
\par\end{centering}

\caption{(Color online) (a) Density profile (solid line) and superfluid density
profile (dashed line) of a trapped 2D Bose gas with $g=0.05$ at $T=0.45T_{F}$,
in unit of the peak density $n_{F}$ of a 2D ideal Fermi gas at $T=0$.
The inset shows the temperature dependence of the relative amplitude
between density and temperature fluctuations of second sound in a
uniform 2D Bose gas. (b) Density fluctuations (in arbitrary unit)
of the lowest three two-fluid modes at the frequencies $A$, $B$
and $C$, as indicated in Fig. \ref{fig2}.}

\label{fig3} 
\end{figure}

The strong coupling between first and second sound could lead to significant
density fluctuations of second sound modes in harmonic traps, as shown
in Fig. \ref{fig3}(b) for the lowest-two first sound ($A$, $B$)
and the lowest second sound modes ($C$) near the BKT transition at
$T=0.45T_{F}\simeq0.9T_{c}$. Surprisingly, the second sound mode
gives much stronger density fluctuation than the first sound mode,
implying that second sound is actually easier to excite and observe
than the first sound. This finding, however, seems consistent with
the earlier observation in the uniform case that, the relative amplitude
between density and temperature fluctuations of second sounds - roughly
given by the LP parameter - is very large near transition, as shown
in the inset of Fig. \ref{fig3}(a).

\begin{figure}
\begin{centering}
\includegraphics[clip,width=0.45\textwidth]{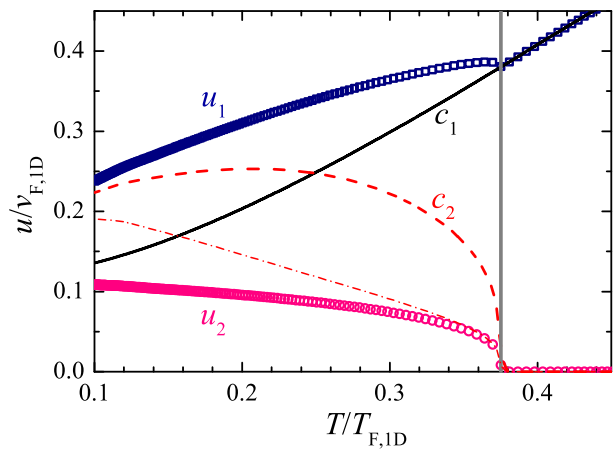} 
\par\end{centering}

\caption{(Color online) Temperature dependence of the 1D first and second sound
velocities (symbols), in unit of the Fermi velocity $v_{F}^{1D}=\sqrt{2k_{B}T_{F}^{1D}/m}$,
where $k_{B}T_{F}^{1D}=(3\pi\hbar\omega_{y}/2)^{2/3}(\hbar^{2}n_{1}^{2}/2m)^{1/3}$
is the characteristic Fermi energy in quasi 1D and $n_{1}$ is the
linear density. The solid and dashed lines show the decoupled first
and second sound velocities, $c_{1}$ and $c_{2}$, respectively.
The dot-dashed line is the second sound velocity with the leading-order
correction in the sound coupling $c_{2}/\sqrt{\gamma}$.}

\label{fig4} 
\end{figure}

We now turn to consider a highly anisotropic harmonic trap with $\omega_{x}\ll\omega_{y}$.
We assume that the number of atoms is large enough so the system is
still in the 2D regime where the local density approximation is applicable.
However, its hydrodynamic behavior is strongly affected by the tight
confinement in the \textit{y}-axis. As discussed by Stringari and
co-workers \cite{Bertaina2010,Hou2013}, Landau's two-fluid hydrodynamic
action $S$ could reduce to a simplified 1D form, due to nonzero viscosity
and thermal conductivity, which lead to invariant local fluctuations
in temperature ($\delta T$) and chemical potential ($\delta\mu$)
as a function of $y$ for any low-energy modes with frequency $\omega\sim\omega_{x}\ll\omega_{y}$.
In other words, we can integrate out the \textit{y} coordinate in
all the thermodynamic variables that enter the hydrodynamic action.
More explicitly, we have a reduced Gibbs-Duhem relation, $\delta P_{1}=s_{1}\delta T+n_{1}\delta\mu$,
where the variables $P_{1}=\int dyP(x,y)$, $s_{1}=\int dys(x,y)$
and $n_{1}=\int dyn(x,y)$ are the \textit{y}-integrals of their 2D
counterparts. All the 1D thermodynamic variables in the simplified
1D hydrodynamic action, except the superfluid density, can be derived
from the reduced Gibbs-Duhem relation using the standard thermodynamic
relations. Hydrodynamic modes in this quasi-1D configuration can then
be solved using the same variational technique as in 2D.

In Fig. \ref{fig4}, we report the 1D first and second sound velocities
for the case of a very weak trapping potential $\omega_{x}\sim0$.
This case is of particular interest since the propagation of sound
waves can be directly observed through the density measurement. Indeed,
both the sound velocities of first and second sound have been recently
measured for a unitary Fermi gas in the similar quasi-1D configuration
\cite{Sidorenkov2013}. In contrast to the uniform 2D case, we find
that the first and second sound velocities no longer exhibit a discontinuity
across the BKT transition. However, there is an apparent kink in the
first sound velocity at transition. This is because after the integral
over the \textit{y}-coordinate, the 1D superfluid density $n_{s1}=\int dyn_{s}(x,y)$
now vanishes as $(T-T_{c})^{1/2}$, approaching to the superfluid
and normal fluid interface. As a result, the correction to the first
sound velocity due to the sound coupling is given by $\Delta c_{1}\simeq\epsilon_{\textrm{LP}}c_{2}^{2}/[2(\epsilon_{\textrm{LP}}+1)c_{1}]\sim(T-T_{c})$
\cite{Hu2010}, which changes the slope of the velocity. We note that
in the quasi-1D configuration, the coupling between first and second
sound is again very strong, as indicated by $\epsilon_{\textrm{LP}}\simeq8$
at the BKT transition (not shown in Fig. \ref{fig4}).

The discretized mode frequency of first and second sound in the presence
of a trapping frequency $\omega_{x}\neq0$ can also be measured experimentally
\cite{Tey2013}. In this case, our calculations predict the similar
pattern for mode frequencies as in Fig. \ref{fig2}. The second sound
mode frequency diverges slower towards the BKT transition, $\omega_{2}\propto\left(T-T_{c}\right)^{-1/4}$,
due to the critical behavior $(T-T_{c})^{1/2}$ of the reduced 1D
superfluid density.

In conclusion, we have presented in this Letter various aspects of
hydrodynamic modes of a 2D dilute Bose gas in harmonic traps. Differently
from the superfluid helium and unitary Fermi gas, first and second
sound have been found to strongly couple with each other. As a consequence,
the second sound has a significant density fluctuation, whose relative
amplitude is much larger than the temperature fluctuation. This makes
the second sound much easier to excite and observe than in a unitary
Fermi gas. We have predicted that the universal jump of superfluid
density at the BKT transition leads to two peculiar features: (1)
the breathing second sound mode frequency in an isotropic trap diverges
like $(T-T_{c})^{-1/2}$ and (2) the first sound velocity of a wave
propagation in a highly anisotropic trap exhibit an apparent kink
right at transition. The observation of these two features gives a
strong evidence of the BKT physics. Our results apply as well to a
2D interacting Fermi gas, which has been recently realized in the
cold-atom laboratory \cite{Martiyanov2010,Frohlich2011,Orel2011}. 
\begin{acknowledgments}
This research is supported by the ARC Discovery Projects (Grant Nos.
FT130100815, DP140103231 and DP140100637) and NFRP-China (Grant No.
2011CB921502).\end{acknowledgments}

\end{document}